\documentclass[aps,11pt,prd,longbibliography]{revtex4-1}
\usepackage{feynmp-auto}
\usepackage{amsmath}
\usepackage[utf8]{inputenc}

\usepackage{bbm}
\usepackage{soul}	
\usepackage{xcolor}

\usepackage{epsfig}
\usepackage{color}
\usepackage{slashed}
\usepackage{comment}
\usepackage{epstopdf}
\usepackage{array}
\usepackage{booktabs}
\usepackage{mathrsfs}
\usepackage[euler]{textgreek}
\usepackage{amsmath}
\usepackage{amsthm}
\usepackage{amsfonts}
\usepackage{amssymb}
\usepackage{graphicx}
\graphicspath{ {Figures/} }
\usepackage[font={small}]{caption}
\usepackage{subcaption}
\usepackage{float}
\usepackage{multirow}
\usepackage[export]{adjustbox}
\usepackage[hang, flushmargin,bottom]{footmisc} 
\usepackage{hyperref}
\hypersetup{
     colorlinks   = true,
     citecolor    = blue
}
\usepackage{placeins}
\usepackage{enumitem}
\usepackage{slashed}
\usepackage{bbm}
\usepackage{appendix}

\usepackage{titlesec}
\titleformat{\chapter}[display]
  {\normalfont\LARGE\bfseries}
  {\chaptertitlename\ \thechapter}{5pt}{\LARGE}
  \titlespacing*{\chapter}{0pt}{-20pt}{35pt}
\usepackage{bigstrut}
\usepackage{pst-node}
\usepackage{pstricks}
\usepackage{physics}
\usepackage{mathtools}
\usepackage{setspace}
\usepackage{fancyhdr}
\usepackage[makeroom]{cancel}
\usepackage{feyn}
\newcommand{\be}{\begin{equation}}
\newcommand{\ee}{\end{equation}}
\newcommand{\bes}{\begin{equation*}}
\newcommand{\ees}{\end{equation*}}

\newcommand{\e}{\text{e}}

\usepackage{hyperref}
\hypersetup{%
  colorlinks = true,
  linkcolor  = blue
}
\usepackage{soul}

\newcommand{\beq}{\begin{equation}}
\newcommand{\eeq}{\end{equation}}

\newcommand{\myComment}[1]{}

\begin{document}
\title{\Large{{Low Scale Seesaw with Local Lepton Number}}}

\author{Hridoy Debnath, Pavel Fileviez P\'erez}
\affiliation{Physics Department and Center for Education and Research in Cosmology and Astrophysics (CERCA), Case Western Reserve University, Cleveland, OH 44106, USA}
\email{hxd253@case.edu, pxf112@case.edu}
\begin{abstract}
We discuss a class of theories for Majorana neutrinos where the total lepton number is a local gauge symmetry. These theories predict a dark matter candidate from anomaly cancellation. We discuss the properties of the dark matter candidate and using the cosmological bounds, we obtain the upper bound on the lepton number symmetry breaking scale. The dark matter candidate has unique annihilation channels due to the fact that the theory predicts a light pseudo-Goldstone boson, the Majoron, and one can obtain the correct relic density in a large fraction of the parameter space.
In this context, the seesaw scale is below the ${\cal{O}}(10^2)$TeV scale and one can hope to test the origin of neutrino masses at current or future colliders. We discuss the lepton number violating Higgs decays and the possibility to observe lepton number violation at the Large Hadron Collider.
\end{abstract}
\maketitle
\section{INTRODUCTION}
%
The origin of the charged fermion and gauge bosons masses in the Standard Model (SM) of particle physics is well-understood. After the discovery of the Brout-Englert-Higgs (BEH) boson at the Large Hadron Collider we know that spontaneous symmetry breaking plays a fundamental role in nature. It is well-known that the SM does not provide a mechanism to generate neutrino masses and one needs to look for a more complete theory.
Today, we know nothing about the nature of neutrinos, they could be Majorana or Dirac fermions. In the case of Majorana neutrinos the total lepton number ($\ell=\ell_e + \ell_\mu + \ell_\tau$ in our notation) symmetry has to be broken in two units, while in the Dirac case the total lepton number is conserved or broken in more than two units.

One of the leading ideas for the origin of neutrino masses is related to the canonical seesaw mechanism~\cite{Minkowski:1977sc,Gell-Mann:1979vob,Mohapatra:1979ia, Yanagida:1979as}, where the SM is extended including at least two copies of right-handed neutrinos. In this context, the SM neutrino masses are suppressed by a new mass scale: $M_\nu \approx  Y_\nu^T M_N^{-1} Y_\nu v_0^2 / 2$,    
where $Y_\nu$ is the Yukawa coupling between the SM neutrinos and the new right-handed neutrinos, $M_N$ defines a new Majorana mass scale and $v_0=246$ GeV is the constant value of the BEH field. This simple mechanism does not predict the scale where the new mass scale $M_N$ is generated, it could be $M_N \approx 10^{14-15}$ GeV if $Y_\nu \sim 1$. If the origin of neutrino masses is related to the so-called canonical scale, $M_{\rm seesaw} \sim 10^{14}$ GeV, there is no way to have direct access to the mechanism of Majorana neutrino masses at current or future collider experiments. 

The scale for total lepton number violation could be around of the TeV scale if $Y_\nu \sim 10^{-6}$ and one could hope to test the origin of Majorana neutrino masses at colliders.
In this case one can look for lepton number violating (LNV) signatures with same sign leptons and multijets as pointed out in Ref.~\cite{Keung:1983uu}. See Ref.~\cite{Cai:2017mow} for a detailed discussion about lepton number violation at colliders and Ref.~\cite{FileviezPerez:2022ypk} for a discussion of different ideas in physics beyond the Standard Model that could predict lepton number violation.

The total lepton number is a key symmetry to understand the origin of neutrino masses. In the SM the total lepton symmetry is a conserved global symmetry at the classical level but broken by the $SU(2)_L$ instantons in three units. One can promote the total lepton number to a local gauge symmetry to understand the link between spontaneous symmetry breaking and neutrino masses. In Refs.~\cite{FileviezPerez:2011pt,Duerr:2013dza,FileviezPerez:2014lnj} the authors studied simple anomaly free theories based on local $U(1)_\ell$. See also Ref.~\cite{Foot:1989ts} for an earlier discussion. 
The simple theories in Refs.~\cite{Duerr:2013dza,FileviezPerez:2014lnj} predict:
\begin{itemize}
    \item a new sector needed for anomaly cancellation.
    \item a dark matter candidate from anomaly cancellation.
    \item the total lepton number must be broken below the multi-TeV scale.
    \item possible lepton number violating signatures at colliders.
\end{itemize}
In this article, we study for the first time a simple theory based on local total lepton number with Majorana neutrinos. We discuss in detail the properties of the dark matter candidate discussing each annihilation channel. This theory predicts the existence of a pseudo-Nambu-Goldstone boson (Majoron) associated to the spontaneous 
breaking of lepton number~\cite{Chikashige:1980ui}. We show that the dark matter candidate can annihilate through different channels containing the Majoron and one can satisfy the relic density constraints in different regions of the parameter space compared to the study in Ref.~\cite{FileviezPerez:2019cyn}. Using the cosmological bounds on the dark matter relic density we point out the upper bound 
on the mass of the new neutral gauge boson associated to the total lepton number gauge symmetry. 

Since the new symmetry breaking scale is below the ${\cal{O}}(10^2)$TeV scale, we discuss the possibility to observe lepton number violation from Higgs decays at the Large Hadron Collider. In the case of the SM-like Higgs one can have a large number of LNV events because the branching ratio, ${\rm BR}(h \to NN)$, could be relative large.
We also show the predictions for the LNV signatures from the other Higgs decays. In this case, one cannot have many events at the LHC due to the fact that the production cross section is small 
if one uses the bounds on the mixing between the SM-like Higgs and the new Higgses coming from direct detection dark matter experiments. The theory studied in this article provides an interesting correlation between the origin of neutrino masses, the nature of dark matter and possible exotic signatures at colliders.  

This article is organized as follows: In Sec.~\ref{LN} we discuss a simple theory for Majorana neutrino masses where the total lepton number is a local gauge symmetry spontaneously broken at the low scale. In Sec.~\ref{DM} we discuss the main properties of the fermionic dark matter candidate predicted from anomaly cancellation in the theory. The lepton number violating Higgs decays at the Large Hadron Collider are discussed in Sec.~\ref{LNV-LHC}, and we summarize our main findings in Sec.~\ref{Summary}.
\section{LEPTON NUMBER AS A LOCAL GAUGE SYMMETRY}
\label{LN}
We can have a simple theory for massive neutrinos based on the gauge symmetry~\cite{FileviezPerez:2011pt,Duerr:2013dza,FileviezPerez:2014lnj} 
$$SU(3)_C \otimes SU(2)_L \otimes U(1)_Y \otimes U(1)_{\ell},$$ 
where $U(1)_\ell$ is defined as the local total lepton number.
It is well-known that the total lepton number is a symmetry conserved 
in the Standard Model at the classical level but broken by three units 
at the quantum level by $SU(2)_L$ instantons. In order to define an anomaly-free theory based on the above gauge symmetry, we need to add new fermions together with right-handed neutrinos ($\nu _R$). In Table.~\ref{1} we list the extra fermions needed for anomaly cancellation as proposed in Ref.~\cite{Duerr:2013dza}. See also the other possibilities in Refs.~\cite{FileviezPerez:2011pt,FileviezPerez:2014lnj}.
\begin{table}[h]
\begin{tabular}{||c c c c c||} 
 \hline
 \hline
 Fields & $SU(3)_C$ & $SU(2)_L$ &$ U(1)_Y$  &$U(1)_\ell$\\ [1.5 ex]
 \hline\hline
 $\Psi_L = \begin{pmatrix}
 \Psi_L^0\\
 \Psi_L^-
 \end{pmatrix}$ &$ \mathbf{1} $& $ \mathbf{2} $ & $-\frac{1}{2}$ & $\ell_1$  \\ [2ex]
 \hline
 $\Psi_R = \begin{pmatrix}
 \Psi_R^0\\
 \Psi_R^-
 \end{pmatrix}$ & $ \mathbf{1} $ & $\mathbf{2}$ & $-\frac{1}{2}$ & $\ell_2$ \\
 \hline
 $\eta_R$ & $ \mathbf{1} $ & $ \mathbf{1} $ & $-1$ & $\ell_1$ \\[2ex]
 \hline
 $\eta_L$ & $ \mathbf{1} $ & $ \mathbf{1} $ & $-1$ & $\ell_2$ \\[2ex]
 \hline
 $\chi_R$ & $ \mathbf{1} $ & $ \mathbf{1} $ & 0 & $\ell_1$\\  [2ex]
 \hline
 $\chi_L$& $\mathbf{1} $ & $\mathbf{1} $ & 0 & $\ell_2$\\[2 ex]
 \hline
\end{tabular}
\caption{\label{1} Fermions needed for anomaly cancellation with $\ell_1 - \ell_2  = - 3$~\cite{Duerr:2013dza}.}
\end{table}
\\
The Lagrangian of this theory is given by 
\begin{align}
\mathcal{L}= & \mathcal{L}_{SM} + i \bar{\nu}_R \slashed \partial \nu_R - g_\ell (\bar{\ell}_L\gamma^{\mu} \ell_L + \bar{e}_R \gamma^{\mu}e_R+ \bar{\nu}_R \gamma^{\mu}\nu_R)Z_{\mu}^\ell \nonumber\\
- & \frac{1}{4}Z_{\mu \nu}^\ell Z^{\ell,\mu \nu}+ \mathcal{L}_{K}^L +\mathcal{L}_{Y}^L -V(H,S) .
\label{eq:LB}
\end{align}
Here $\ell_L \sim (\mathbf{1},\mathbf{2},-1/2,1)$ and $e_R \sim (\mathbf{1},\mathbf{1},-1,1)$ are the SM leptonic fields, $\mathcal{L}_{SM}$ is the SM Lagrangian and $V(H,S)$ contains the new terms in the scalar potential.
Here we neglect the kinetic mixing term between the two Abelian gauge symmetries for simplicity.
The kinetic terms for the new fields can be written as
\begin{align}
    \mathcal{L}_K^L & =  i\bar{\Psi}_L\slashed D\Psi_L+i\bar{\Psi}_R\slashed D\Psi_R+ i \bar{\eta}_L \slashed D \eta_L+i\bar{\eta}_R \slashed D \eta_R\nonumber\\ 
    & +i \bar{\chi}_L \slashed D \chi_L+i \bar{\chi}_R \slashed D \chi_R + (D_{\mu}S)^{\dagger}(D^{\mu}S),
    \label{eq:LB}
\end{align}
and the new Yukawa interactions read as 
\begin{eqnarray}
-\mathcal{L}_Y^L &=&  y_1 \bar{\Psi}_LH\eta_R + y_2 \bar{\Psi}_R H\eta_L +  y_3 \bar{\Psi}_L\tilde{H}\chi_R+y_4\bar{\Psi}_R\tilde{H}\chi_L  \nonumber \\
&+& y_{\Psi} \bar{\Psi}_L\Psi_R S^*
+ y_{\eta}\bar{\eta}_R \eta_L S^* + y_{\chi}\bar{\chi}_R \chi_L S^* \ + \ {\rm{h.c.}}.
    \label{eq:LB}
\end{eqnarray}
Here $H \sim (\mathbf{1},\mathbf{2},1/2,0)$ is the Standard Model Higgs, and $\tilde{H}=i\sigma_2 H^*$. 
Here we do not consider the case when $\ell_1 \neq - \ell_2$ and assume $\ell_i \neq \pm 1$. Notice that when we have fractional values for $\ell_i$ one makes sure that higher-dimensional operators do not affect the stability of our dark matter candidate. 
The case $\ell_1=-\ell_2=-3/2$ is interesting but less generic.
In this case the Yukawa interactions $\chi_L \chi_L S^*$ and $\chi_R \chi_R S$ are allowed by the gauge symmetry, and the dark matter candidate is a Majorana fermion. In this article we focus on the most generic cases with $\ell_1-\ell_2=-3$,  $\ell_1 \neq -\ell_2$ and $\ell_i \neq \pm 1$. Notice that the condition $\ell_i \neq \pm 1$ is needed to avoid the mixing between the $\chi_L$, $\chi_R$ and the right-handed neutrinos.

The new Higgs quantum numbers are determined by anomaly cancellation condition, $\ell_1 - \ell_2  = - 3$, and the above Yukawa interactions. Therefore, the new Higgs transforms as:
\begin{equation}
S \sim (\mathbf{1},\mathbf{1},0,3).
\end{equation}
For some studies in this context see Refs.~\cite{FileviezPerez:2011pt,Schwaller:2013hqa,FileviezPerez:2019cyn,Carena:2022qpf,Madge:2018gfl,FileviezPerez:2015mlm}.

In this theory we can generate Majorana neutrinos using the following interactions
\begin{eqnarray}
    -\mathcal{L}_{\nu} \supset  Y_{\nu} \ \bar{\ell}_L\tilde{H}\nu_{R} + \lambda_R \ \nu_R^T C \phi \nu_R \ + \ {\rm h.c.}.
\end{eqnarray}
Here $\phi\sim(\mathbf{1},\mathbf{1},0,-2)$ is a new Higgs field needed to implement the seesaw mechanism. After the local lepton number symmetry, $U(1)_{\ell}$, is broken the theory has an accidental global $U(1)$ symmetry:
\begin{eqnarray}
 \Psi_L \to e^{i\theta} \Psi_L, \ && \Psi_R \to e^{i \theta} \Psi_R, \nonumber \\
 \eta_L \to e^{i\theta} \eta_L, \  && \eta_R \to e^{i\theta}  \eta_R, \nonumber \\
 \chi_L \to e^{i\theta}  \chi_L, \ && \chi_R \to e^{i\theta} \chi_R. \nonumber
\end{eqnarray}
Therefore, the lightest field in the new sector is stable. In order to have a consistent scenario for cosmology we assume that the lightest new stable field is the neutral Dirac field: 
$\chi=\chi_L+\chi_R$. Therefore, one can say that this theory predicts a cold dark matter candidate from anomaly cancellation and its stability is a natural consequence from spontaneous symmetry breaking.
Notice that field $\Psi=\Psi_L^0 + \Psi_R^0$ is ruled out as a dark matter candidate because it has a large coupling to the $Z$ gauge boson and one cannot satisfy the dark matter direct detection constraints.

 {{Higgs Sector}}:
 The Higgs sector is composed of the SM Higgs, $H$, and the new Higgses, $S$ and $\phi$.
 The scalar potential in this theory is given by  
 \begin{eqnarray}
 V(H,S,\phi)&=&-m_H^2 H ^{\dagger}H+\lambda(H^{\dagger}H)^2-m_s^2 S ^{\dagger}S + \lambda_s (S^{\dagger}S)^2-m_{\phi}^2 \phi ^{\dagger}\phi \nonumber
 \\
 &+& \lambda_{\phi}(\phi^{\dagger}\phi)^2+\lambda_1(H^{\dagger}H)S^{\dagger}S + \lambda_2(H^{\dagger}H)\phi^{\dagger}\phi+ \lambda_3(S^{\dagger}S)\phi^{\dagger}\phi.
 \end{eqnarray}
Notice that this scalar potential has the global symmetry 
$O(4)_H \otimes U(1)_\phi \otimes U(1)_S $. 
The scalar fields in this theory can be written as 
\begin{eqnarray}
H&=&\begin{pmatrix}
h^+\\
\frac{1}{\sqrt{2}}(v_{0} + h_0) \e^{i \sigma_0/v_0} 
\end{pmatrix}, \\
S &=& \frac{1}{\sqrt{2}}\left(v_{s} + h_s \right) \e^{i \sigma_s/v_s}, 
\end{eqnarray}
and
\begin{eqnarray}
\phi &=& \frac{1}{\sqrt{2}}\left( v_{\phi} + h_{\phi} \right) \e^{i \sigma_{\phi}/v_{\phi}} .
\end{eqnarray}
After the spontaneous symmetry breaking, the constant part of the scalar potential can be written as
\begin{eqnarray}
V(v_0,v_{L},v_{\phi})&=&-\frac{1}{2}m_H^2v_0^2+\frac{\lambda}{4}v_0^4-\frac{1}{2} m_s^2v_s^2+\frac{\lambda_s}{4}v_s^4-\frac{1}{2}m_{\phi}^2v_{\phi}^2
\nonumber\\
&+& \frac{\lambda_{\phi}}{4}v_{\phi}^4 +\frac{\lambda_1}{4}v_0^2v_s^2+\frac{\lambda_2}{4}v_0^2v_{\phi}^2+\frac{\lambda_3}{4}v_{\phi}^2v_s^2,
\end{eqnarray}
and the minimization conditions read as
\begin{eqnarray}
-m_H^2v_0+\lambda v_0^3+\frac{1}{2}\lambda_1 v_s^2v_0+\frac{1}{2}\lambda_2 v_{\phi}^2v_0 = 0,
\\
-m_{\phi}^2v_{\phi}+\lambda_{\phi}v_{\phi}^3+\frac{1}{2}\lambda_2v_0^2v_{\phi}+\frac{1}{2}\lambda_3v_s^2v_{\phi}=0,
\end{eqnarray}
and
\begin{eqnarray}
-m_{s}^2v_{s}+\lambda_{s}v_{s}^3+\frac{1}{2}\lambda_1 v_0^2v_{s}+\frac{1}{2}\lambda_3v_{\phi}^2v_{s}=0.
\end{eqnarray}
In this theory, the mass matrix for the CP-even Higgses, in the $(h_0,h_s,h_{\phi})$ basis, can be written as
\begin{eqnarray}
M_{even}^2=\begin{pmatrix}
2\lambda v_0^2 & \lambda_1 v_0 v_s &\lambda_2 v_0 v_{\phi} \\
\lambda_1 v_0 v_s & 2\lambda_s v_s^2 & \lambda_3 v_sv_{\phi}\\
\lambda_2 v_0 v_{\phi} & \lambda_3 v_sv_{\phi} & 2 v_{\phi}^2\lambda_{\phi}
\end{pmatrix}.
\end{eqnarray}
In our notation the physical Higgses, $(h,H_1,H_2)$, are defined as
\begin{eqnarray}
\begin{pmatrix}
h_0\\
h_s\\
h_{\phi} 
\end{pmatrix} =
U
\begin{pmatrix}
h\\
H_1\\
H_2
\end{pmatrix}.
\label{Umixing}
\end{eqnarray}
There are three CP-odd Higgses in this theory, two of them are Goldstone's bosons eaten by the neutral gauge bosons. Notice that the CP-odd Higgses, $\sigma_s$ and $\sigma_\phi$, masses are protected by the a shift symmetry: $\sigma_i \to \sigma_i + c$, where $c$ is a constant. This shift symmetry is broken by the dimensional five 
term in the scalar potential:
\begin{eqnarray}
 V(H,S,\phi) \supset  \lambda_{M} \frac{S^2 \phi^3}{\Lambda}+ {\rm h.c.}. 
 \end{eqnarray}
This term is allowed by all the symmetries of the theory and tells us that there is only one Nambu-Golstone boson in the new sector.  
Notice that this term breaks the global symmetry, $U(1)_\phi \otimes U(1)_S $, to a new $U(1)$ symmetry.
The CP-odd mass matrix in the $(\sigma_s ,\sigma_{\phi})$ basis can be written as
\begin{eqnarray}
M_{odd}^2 = \frac{\lambda_M}{\sqrt{2}\Lambda}\begin{pmatrix}
 2 v_{\phi}^3 & 3 v_s v_{\phi}^2\\
 3 v_s v_{\phi}^2 & \frac{9}{2} v_s^2 v_{\phi}
 \end{pmatrix},
\end{eqnarray}
with eigenvalues
\begin{eqnarray}
M^2_{G_{\ell}} =0  \hspace{0.2 cm}\text{and  } M^2_{J}= \frac{\lambda_M v_{\phi}}{2\sqrt{2}\Lambda}\left(4v_{\phi}^2 + 9 v_s^2 \right).
\label{Majoron}
\end{eqnarray}
In this theory, the $Z_{\ell}$ mass can be written as 
\begin{eqnarray}
    M_{Z_{\ell}}^2=g_{\ell}^2 \left( 9 v_s^2 \ + \ 4 v_{\phi}^2 \right).
    \label{Zell}
\end{eqnarray}
Notice that the mass of the leptophilic gauge boson can be used as the seesaw scale in this theory because it tells us about the scale where lepton number is spontaneously broken.

Using Eqs.(\ref{Majoron}) and (\ref{Zell}) one can estimate the Majoron mass as a function of the ratio between the new gauge boson mass and gauge coupling: $M_J^2=\lambda_M M_{Z_\ell}^3 \cos \beta / (4 \sqrt{2} g_\ell^3 \Lambda)$.
One finds that $M_J \sim 10^{-4}$ GeV when $\lambda_M \sim 1$, $M_{Z_\ell}/g_\ell \sim 10$ TeV, $\cos \beta \sim 1$ and $\Lambda \sim M_{Pl}$, with $M_{Pl}$ being the Planck scale. As we will discuss, in our case the seesaw scale will be below ${\cal{O}} (10^2)$ TeV. Therefore, the Majoron will decay fast into neutrinos and it cannot be a dark matter candidate. See for example Ref.~\cite{Heeck:2019guh} for a detailed discussion of the Majoron properties. For the cosmological bounds on the Majoron interactions with neutrinos, see for example Refs.~\cite{Escudero:2019gvw,Sandner:2023ptm}. 

The CP-odd Higgs eigenstates are defined by
\begin{eqnarray}
\begin{pmatrix}
\sigma_s \\
\sigma_{\phi} 
\end{pmatrix} =
\begin{pmatrix}
\cos \beta & \sin \beta \\
- \sin \beta & \cos \beta \\
\end{pmatrix}
\begin{pmatrix}
G_{\ell} \\
J
\end{pmatrix},
\end{eqnarray}
where 
\begin{equation}
\tan 2 \beta= \frac{12 v_s v_\phi}{4 v_\phi^2 - 9 v_s^2}.
\end{equation}
\newpage
Fermionic Fields:
\begin{itemize}
\item Neutral Dirac Fermions:
%
The mass matrix for the neutral fermions can be written, in the  $(\chi_{L}^0 \quad \Psi_L^0) $ and $ (\chi_{R}^0 \quad \Psi_R^0)$ basis, as
\begin{eqnarray}
    -\mathcal{L}\supset (\overline{\chi_{R}^0} \quad \overline{\Psi_R^0}) \mathcal{M}_0  \begin{pmatrix}
 \chi_L^0\\
 \Psi_L^0
 \end{pmatrix} + {\rm{h.c.}},
\end{eqnarray}
where 
\begin{eqnarray}
    \mathcal{M}_0 = \frac{1}{\sqrt{2}} \begin{pmatrix}
y_{\chi}v_s  & y_3 v_0 \\
 y_4v_0 & y_{\Psi}v_s
 \end{pmatrix}.
\end{eqnarray}
One can diagonalize the mass matrix as follows 
\begin{eqnarray}
     \mathcal{M}_0^{diag}= N_R^{\dagger} \mathcal{M}_0 N_L.
\end{eqnarray}
where the neutral fields are related by the $N_L $ and $N_R$ mixing matrices as given by
\begin{eqnarray}
\begin{pmatrix}
 \chi_L^0 \\
 \Psi_L^0
 \end{pmatrix} = N_L \begin{pmatrix}
 \chi_{1L}^0\\
 \chi_{2L}^0
 \end{pmatrix} , 
 \ {\rm{and}} \
 \begin{pmatrix}
 \chi_R^0 \\
 \Psi_R^0
 \end{pmatrix} =  N_R\begin{pmatrix}
 \chi_{1R}^0\\
 \chi_{2R}^0
 \end{pmatrix}.
 \end{eqnarray} 
 In the limit when, $v_s\gg v_0$, the mass of the dark matter candidate is given $M_\chi= y_\chi v_s / \sqrt{2}$.
 \item Charged Fermions: 
 The mass matrix for the charged fermions can be written, in the  $(\eta_{R}^- \quad \Psi_R^-) $ and $ (\eta_{L}^- \quad \Psi_L^-)$ basis, as
  \begin{eqnarray}
    -\mathcal{L}\supset (\overline{\eta_{R}^-} \quad \overline{\Psi_R^-}) \mathcal{M}_C  \begin{pmatrix}
 \eta_L^-\\
 \Psi_L^-
 \end{pmatrix} + {\rm{h.c.}},
\end{eqnarray}
where \begin{eqnarray}
    \mathcal{M}_C = \frac{1}{\sqrt{2}} \begin{pmatrix}
y_{\eta}v_s  & y_1 v_0 \\
 y_2v_0 & y_{\Psi}v_s
 \end{pmatrix}.
\end{eqnarray}
 In our notation, the mass matrix can be diagonalized as follows 
\begin{eqnarray}
     \mathcal{M}_C^{diag}= V_R^{\dagger}\mathcal{M}_C V_L.
\end{eqnarray}
 while the physical charged fields are related by the $V_L$ and $V_R$ matrices 
 \begin{eqnarray}
\begin{pmatrix}
 \eta_L^- \\
 \Psi_L^-
 \end{pmatrix} = V_L \begin{pmatrix}
 \chi_{1L}^-\\
 \chi_{2L}^-
 \end{pmatrix} , 
 \ {\rm{and}} \
 \begin{pmatrix}
 \eta_R^- \\
 \Psi_R^-
 \end{pmatrix} =  V_R\begin{pmatrix}
 \eta_{1R}^-\\
 \chi_{2R}^-
 \end{pmatrix}.
 \end{eqnarray}
\item Neutrino Masses: In this theory, the neutrino masses are generated through the type I seesaw mechanism and the SM neutrino mass matrix is given by
\begin{equation}
    M_\nu = \frac{v_0^2}{2} Y_\nu M_N^{-1} Y_\nu^T,
\end{equation}
where 
\begin{equation}
M_N=\sqrt{2} \lambda_R v_\phi= \frac{\lambda_R}{\sqrt{2}} \frac{M_{Z_\ell}}{g_\ell} \cos \beta.
\end{equation}
Here we used $v_\phi=v \cos \beta / 2$ and $v_s=v \sin \beta / 3$.
Therefore, in this theory the upper bound on the seesaw scale is determined by the ratio $M_{Z_\ell}/ g_\ell$ and the perturbative bound on the Yukawa coupling $\lambda_R$.  
\end{itemize}
\section{DARK MATTER CONSTRAINTS}
\label{DM}
 \begin{figure}[h]
     \centering 
     \begin{subfigure}[h]{0.45\textwidth}
         \centering
         \includegraphics[width=\textwidth]{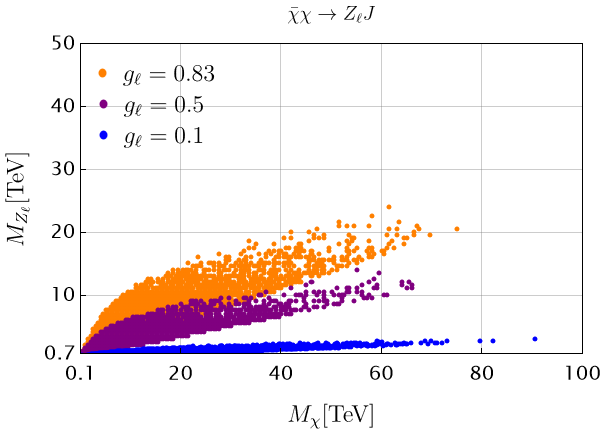 }
     \end{subfigure}
     \hfill
     \begin{subfigure}[h]{0.45\textwidth}
         \centering
         \includegraphics[width=\textwidth]{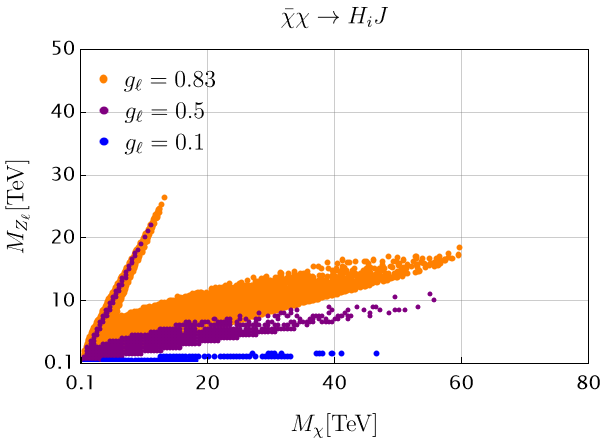}
     \end{subfigure}
        \caption{Allowed regions by the cosmological bounds on dark matter relic density~\cite{Planck:2018vyg} and perturbative bound on the $y_\chi$ coupling. In the left panel, we have the annihilation into $Z_\ell J$, while in the right panel, one has the $H_i J$ channel. We show the results for different values of the gauge coupling $g_\ell$ and assume that the mass of the new Higgses are equal to $1$ TeV.}
        \label{DMDMZJ}
\end{figure}
\begin{figure}[h]
    \centering
  \includegraphics[width=\textwidth]{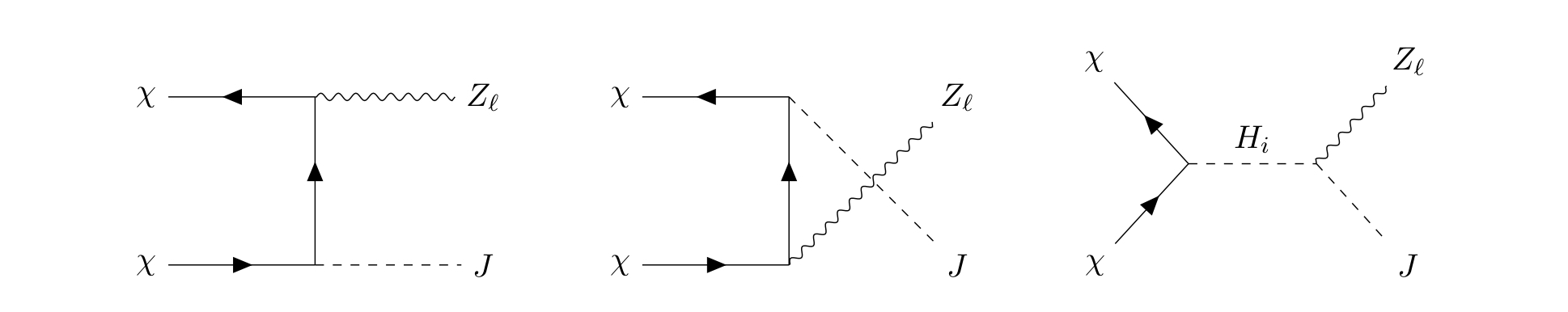}
    \caption{Feynman graphs for the annihilation channel $\bar{\chi} \chi \to Z_\ell J$.}
    \label{ZJgraphs}
\end{figure}
\begin{figure}[h]
     \centering 
     \begin{subfigure}[h]{0.45\textwidth}
         \centering
         \includegraphics[width=\textwidth]{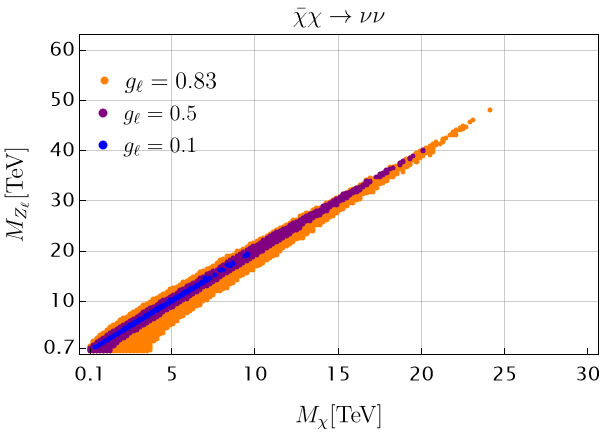 }
     \end{subfigure}
     \hfill
     \begin{subfigure}[h]{0.45\textwidth}
         \centering
         \includegraphics[width=\textwidth]{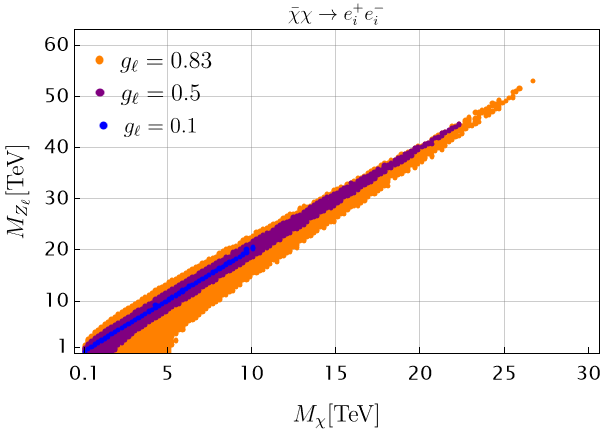}
     \end{subfigure}
     \caption{Allowed regions by the cosmological bounds on dark matter relic density~\cite{Planck:2018vyg} and perturbative bound on the $y_\chi$ coupling, when only the annihilation into leptons is included. In the left panel, we consider the annihilation into neutrinos, while in the right panel, we have the annihilation into charged leptons.}
     \label{DMDMleptons}
\end{figure}
 \begin{figure}    
     \begin{subfigure}[h]{0.45\textwidth}
         \centering
         \includegraphics[width=\textwidth]{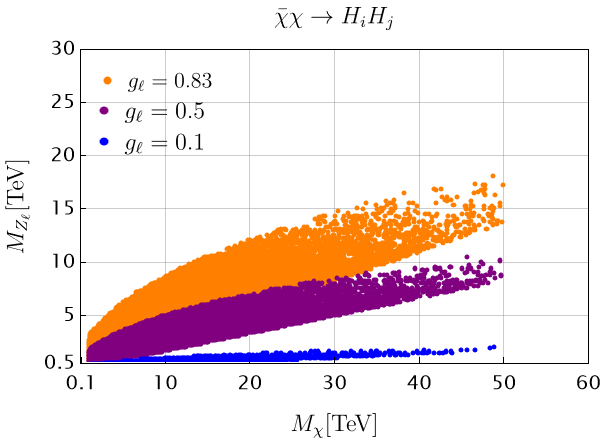} 
     \end{subfigure}
     \hfill
     \begin{subfigure}[h]{0.45\textwidth}
         \centering
         \includegraphics[width=\textwidth]{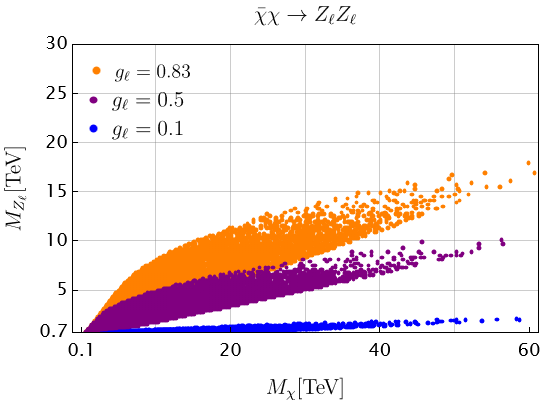}
     \end{subfigure}
     \label{DMDMHH}
    \caption{Allowed regions by the cosmological bounds on dark matter relic density~\cite{Planck:2018vyg} and perturbative bound on the $y_\chi$ coupling, when only the annihilation into two Higgses or two gauge bosons are included. In the left panel we consider the annihilation into two Higgses, while in the right panel we have the annihilation into two gauge bosons. Here we use $M_{H_1}=M_{H_2}= 1$ TeV.} 
    \label{DMDMZLZL}
  \end{figure}
\begin{figure}   
     \begin{subfigure}[h]{0.45\textwidth}
         \centering
         \includegraphics[width=\textwidth]{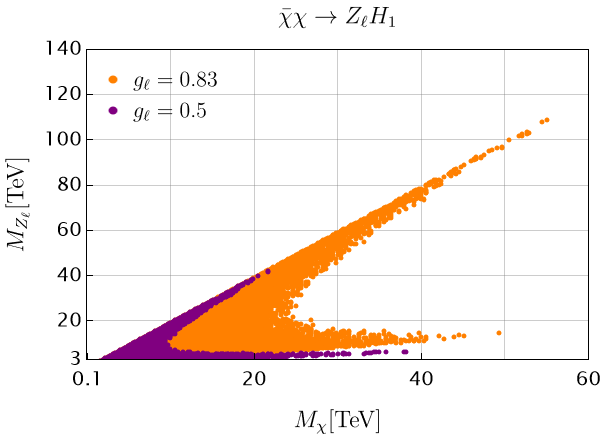}
     \end{subfigure}
     \hfill
     \begin{subfigure}[h]{0.45\textwidth}
         \centering
         \includegraphics[width=\textwidth]{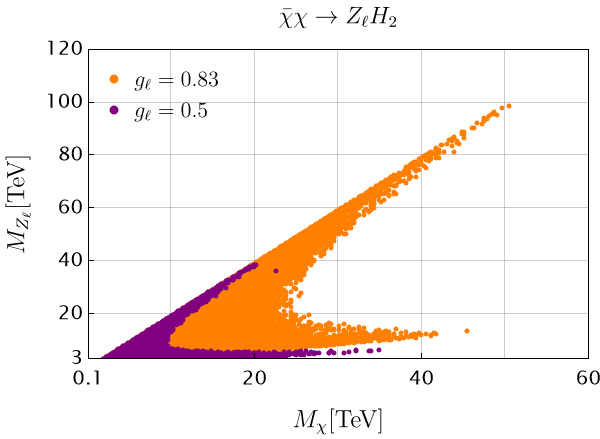}
     \end{subfigure}
 \caption{Allowed regions by the cosmological bounds on dark matter relic density~\cite{Planck:2018vyg} and perturbative bound on the $y_\chi$ coupling, when only the annihilations into a Higgs and a gauge boson are included. In the left panel we have $\bar{\chi} \chi \to Z_\ell H_1$, while in the right panel we show the results for $\bar{\chi} \chi \to Z_\ell H_2$. Here we use $M_{H_1}=M_{H_2}= 1$ TeV.}        
        \label{DMDMZH}
\end{figure}

We have discussed above that this theory predicts a dark matter from anomaly cancellation. In this context, the dark matter is a Dirac fermion, $\chi=\chi_L + \chi_R$. After symmetry breaking $\chi_L$ and $\chi_R$ have the same quantum numbers. The dominant dark matter annihilation channels are 
$$\bar{\chi} \chi \rightarrow e_ie_i,\nu \nu,Z_{\ell}Z_{\ell},H_iH_j,Z_{\ell}H_i, Z_\ell J, J H_i.$$
Notice that the existence of the annihilation channels containing the pseudo-Goldstone boson $J$, the Majoron, is quite unique. These channels will allow us to obtain the correct relic density in regions of the parameter space far from the $Z_\ell$ resonance.

One can compute the relic density using~\cite{Gondolo:1990dk} 
\begin{eqnarray}
    \Omega_{DM}h^2=\frac{1.05\times10^9 \ {\rm{GeV}}^{-1}}{J(x_f)M_{Pl}},
\end{eqnarray}
where the function $J(x_f)$ can be written as 
\begin{eqnarray}
    J(x_f)=\int_{x_f}^{\infty} \frac{g_{*}^{1/2}(x) \left<\sigma v \right>(x)}{x^2} \,dx, 
\end{eqnarray}
where $g_*$ is the total number of degrees of freedom at freeze-out  and the thermal average cross-section times velocity is given by 
\begin{eqnarray}
    \left<\sigma v \right>(x)=\frac{x}{8 M_{\chi}^5K_2^2(x)}\int_{4M_{\chi}^2}^{\infty}\sigma \times(s-4M_{\chi}^2)\sqrt{s}\hspace{0.1 cm}K_1\left(\frac{x\sqrt{s}}{M_{\chi}}\right)  \,ds. 
\end{eqnarray}
Here $x=M_{\chi}/T$,  $K_1(x)$ and $K_2(x)$ are the modified Bessel Functions. The freeze-out parameter $x_f$ can also be computed by using 
\begin{eqnarray}
    x_f=\text{ln}\left(\frac{0.038 \hspace{0.1 cm}g\hspace{0.1 cm} M_{Pl}\hspace{0.1 cm}M_{\chi} \left<\sigma v \right>(x_f)}{\sqrt{g_{*}\hspace{0.1 cm}x_f}}\right).
\end{eqnarray}
where g is the effective number of degrees of freedom of the dark matter particle and $M_{Pl}= 1.22 \times 10^{19}$ GeV. In order to understand the importance of each annihilation channel, we study the contribution of each channel independently, and later taking into account all relevant annihilation channels we find the upper bound on the lepton number symmetry breaking scale.
\begin{figure}   
     \begin{subfigure}[h]{0.45\textwidth}
         \centering
         \includegraphics[width=\textwidth]{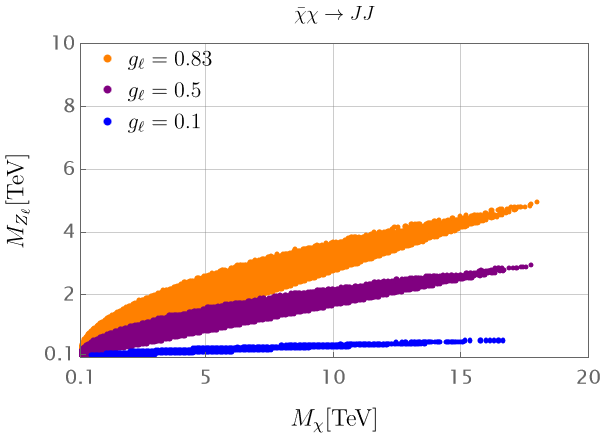}
     \end{subfigure}
     \hfill
     \begin{subfigure}[h]{0.45\textwidth}
         \centering
         \includegraphics[width=\textwidth]{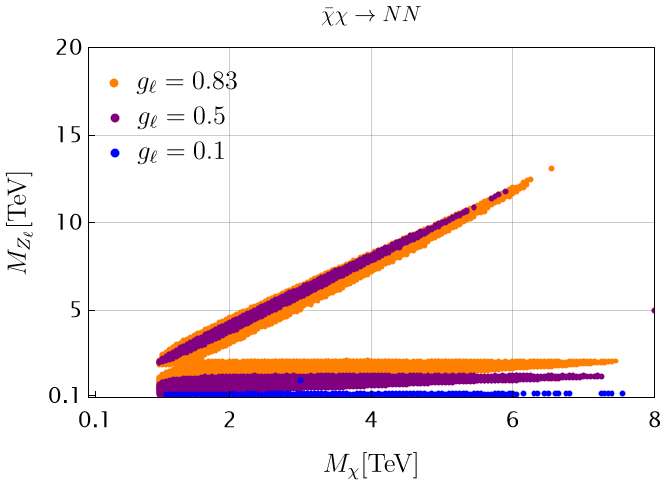}
     \end{subfigure}
 \caption{Allowed region by the cosmological relic density~\cite{Planck:2018vyg} and perturbative bound  on the Yukawa coupling $y_{\chi}$. In the left-panel we show the annihilation to Majorons, while in te right panel we have the annihilation into right-handed neutrinos. For illustration, we used $M_N =1 $ TeV.}        
        \label{DMDMNN}
\end{figure}
\begin{figure}   
         \centering
         \includegraphics[width=0.65\textwidth]{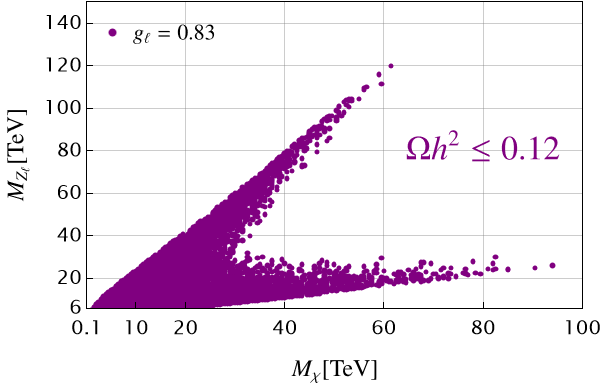}
\caption{Allowed region by the cosmological relic density~\cite{Planck:2018vyg} and perturbative bound  on the Yukawa coupling $y_{\chi}$.  Here we include all relevant annihilation channels. For illustration, we use $M_{H_1}=M_{H_2}= 1$ TeV, and the mixing between the heavy Higgses is ${\theta_H}=\pi/6$.}
    \label{DMDMall}
\end{figure}\\
In this theory, we have unique annihilation channels since the dark matter can annihilate into a light Majoron, $J$. Then, we start our discussion taking into account only the annihilation channel $\bar{\chi} \chi \to Z_\ell J$. This channel has three contributions, see Fig.~\ref{ZJgraphs}.

In Fig.~\ref{DMDMZJ} (left panel) we show the allowed region when only this channel is included in the calculation of the relic density. As one can appreciate, there is an upper bound on the $Z_\ell$ mass around $30$ TeV. This annihilation channel can give us the correct relic density in a large fraction of the parameter space. We show the results for three different values of the gauge coupling, $g_\ell=0.83$ (in orange), $g_\ell=0.5$ (in purple), and $g_\ell=0.1$ (in blue). Notice that when we decrease the value of the gauge coupling $g_\ell$, from 0.83 to 0.1, the allowed region is smaller because the cross section for the process $\chi \bar{\chi} \to Z_\ell J$ is smaller for smaller $g_\ell$ and one obtains the correct relic density only in a reduced region of the parameter space. 
For all numerical examples, we use the values $\ell_1=1/2$ and $\ell_2=7/2$. The main results are very similar if we choose different values for $\ell_1$ and $\ell_2$ charges. This channel is important because one can have a large annihilation cross section when $M_\chi > M_{Z_\ell}/2$ because the Majoron mass is very small and one does need to rely in a resonance to achieve the correct relic density. We show similar results for the annihilation channel $\bar{\chi} \chi \to H_i J$ in the right panel in Fig.~\ref{DMDMZJ}. However, in this case one has also the allowed solutions when you are close to the $Z_\ell$ resonance because one of the processes is $\chi \bar{\chi} \to Z_\ell^* \to H_i J$, and then when $M_\chi \sim M_{Z_\ell}/2$ one can have the large contribution of the $Z_\ell$ resonance. The different colored regions correspond to the different values of the gauge couplings as in the left-panel. The maximal value of $g_\ell$ is determined by the perturbative bound on the $S^\dagger S Z_\ell Z_\ell$ coupling, which gives us that $g_\ell \leq 0.83$.

In Fig.~\ref{DMDMleptons} we show the allowed parameter when we include only the dark matter annihilation into leptons. In the left panel we have the annihilation into neutrinos, while in the right panel we have the annihilation into charged leptons. These channels can give the correct relic density only when one has the $Z_\ell$ resonance, i.e. $M_\chi \sim M_{Z_\ell}/2$. The different colored regions correspond to different values of the gauge coupling. As one expects, when the gauge coupling is smaller it is more difficult to find solutions allowed by the relic density constraints because the annihilation cross section is smaller and one typically obtains too much relic density for smaller couplings. Therefore, when the coupling is smaller one has allowed solutions only when the dark matter mass is very close to $M_{Z_\ell}/2$. 

In Fig.~\ref{DMDMZLZL} we show the same results for the $H_i H_j$ and $Z_\ell Z_\ell$ annihilation channels. When these channels are kinematically allowed, one can have 
a relic density in agreement with experiments in a large fraction of the $M_{Z_\ell}-M_\chi$ plane. The main contributions for the annihilation cross sections for these channels are from the u and t channels. Since the masses of $Z_\ell$ and the new Higgses must be well above the electroweak scale, these processes are mainly relevant when the dark matter mass is above the electroweak scale.

In Fig.~\ref{DMDMZH}, we show the allowed parameter space when one has only the $Z_\ell H_1$ (left panel) or the $Z_\ell H_2$ (right panel) channels. In this case, one has two main regions, the region around the $Z_\ell$ resonance and the region when $2 M_\chi \gg M_{Z_\ell} + M_{H_i}$. 
Here we show the allowed solutions by the relic density constraints only for two values of the gauge couplings, $g_\ell=0.5$ (in orange) and $g_\ell=0.83$ (in purple), because there are no allowed solution when $g_{\ell}=0.1$. Notice that in this case one has two main regions, below the $Z_\ell$ resonance and the $u$ and $t$-channels allow us to obtain solutions far from the resonance. 

In Fig.~\ref{DMDMNN} the allowed $M_{Z_{\ell}}-M_{\chi}$ plane is shown by considering only the $\chi \bar{\chi}\rightarrow JJ$  (left-panel) and $\chi \bar{\chi}\rightarrow NN$ (right-panel) annihilation channels. In this theory, the $\chi \bar{\chi}\rightarrow JJ$  annihilation channel is velocity suppressed, and the allowed parameter space by the relic density doesn't satisfy the collider bounds on the $Z_\ell$ mass. The $\chi \bar{\chi}\rightarrow NN$  channel can satisfy both the relic density and collider bounds only in the resonance region. These channels are clearly unique because they are presence because the theory predicts the existence of the 
Majoron and the right-handed neutrinos. The colored regions correspond to the same values of the gauge couplings as in the previous figures. Notice that in the case when we study the annihilation into two Majorons, one has only the $u$ and $t$ contributions. For the annihilation into two right-handed neutrinos we have also $\chi \bar{\chi}\rightarrow Z_\ell^* \rightarrow NN$, and then we can have also the allowed region close to the $Z_\ell$-resonance.

Finally, in Fig.~\ref{DMDMall} we include all annihilation channels discussed above
and show the allowed parameter space by the relic density constraints. As one expects, there are two main regions: a) The $Z_\ell$-resonance (less generic) and b) the region when the dark matter is far from the $Z_\ell$-resonance and the annihilation channels with the Majoron and the new Higgses are allowed. The second region is more generic because it does not rely in any particular relation of the masses of the fields involved. Clearly, if we think about the most generic allowed solutions the upper bound on the leptophilic gauge boson mass is around 30 TeV.
See appendix~\ref{AppendixB} for the Feynman graphs for each different annihilation channel. The numerical results in Fig.~\ref{DMDMall} are in agreement with the perturbative unitarity bounds~\cite{Griest:1989wd}.
Notice that these results tell us that the seesaw scale in this theory is below the ${\cal{O}}(10^2)$ TeV scale and one can hope to test the origin of neutrino masses at colliders.\\
\begin{figure}[h]
    \centering
    \includegraphics[width=\textwidth]{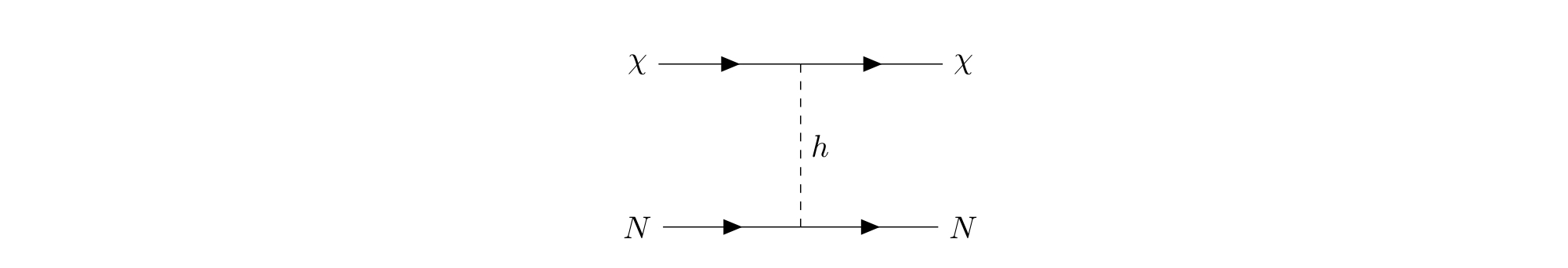}
    \caption{Feynman diagram for the dark matter-nucleon cross-section mediated by the SM-like Higgs boson.}
    \label{DD}
\end{figure}
In this theory, the main contribution to the spin-independent dark matter - nucleon cross section is mediated by the SM-like Higgs as we show in Fig.~\ref{DD}. This cross-section is given by
\begin{eqnarray}
    \sigma_{SI}= \frac{72}{\sqrt{2}} \frac{G_F}{\pi M_h^4} \frac{g_\ell^2 M_\chi^2 \sin^2 \theta}{M_{Z_\ell}^2 \sin^2 \beta} \frac{M_{\chi}^2 m_N^4}{(M_{\chi}+m_N)^2} f_N^2.
\end{eqnarray}
Here $m_N$ is the nucleon mass, $G_F$ is the Fermi constant, and $f_N=0.3$~\cite{Hoferichter:2017olk} is the effective Higgs-nucleon-nucleon coupling. 
Here $\sin \theta=U_{21} U_{11}$, where the $U_{ij}$ are the elements of the mixing matrix in Eq.(\ref{Umixing}). Notice that if we include the kinetic mixing between the two Abelian gauge groups the new neutral gauge boson can also mediate this scattering process. See Ref.~\cite{Kopp:2009et} for the calculation of this cross section at one-loop level. This contribution to the cross section for the dark mater-nucleon scattering is highly suppressed by the mass of the new gauge boson but it is independent of the mixing angle $\theta$. In this theory, the new gauge boson $Z_\ell$ mediates the dark matter-electron scattering but it is very suppressed by the ratio $(g_\ell/M_{Z_\ell})^4$. 
\begin{figure}[h]
     \centering 
     \begin{subfigure}[h]{0.9\textwidth}
         \centering
         \includegraphics[width=\textwidth]{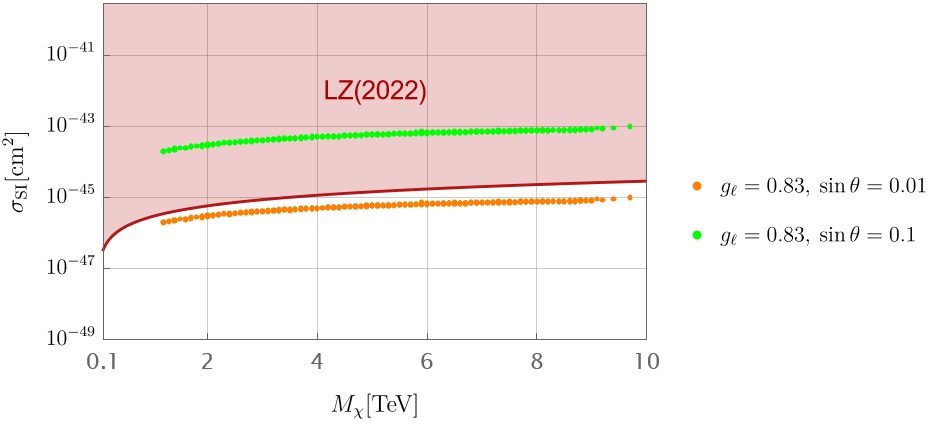}
     \end{subfigure}
     \label{DMdirect}
    \caption{Numerical values for the Spin-independent cross-section for dark matter-nucleon-scattering. The red-shaded region is excluded by the LZ experiment~\cite{LZ:2022ufs}.  For $g_{\ell}$= 0.83, orange and green lines  represent the  spin-independent cross-section for  $\sin{\theta}= 0.01$ and  $\sin{\theta}$= 0.1, respectively. We have used $\sin {\beta}$=0.54 
    (when $v_s=v_\phi$) and $f_N$=0.3.}
    \label{DMSIcrosssection}
\end{figure}\\
In Fig.~\ref{DMSIcrosssection} we show the numerical results for the spin-independent dark matter-nucleon cross section for different values of $g_\ell$ and $\sin \theta$. In the case when $g_{\ell}= 0.83$, the orange and green lines represent the  spin-independent cross section for $\sin {\theta}=0.01$ and  $\sin {\theta}=0.1$, respectively. Notice that all numerical results presented in Fig.~\ref{DMSIcrosssection} are valid for the different values of the input parameters, $g_\ell$, $M_\chi$ and $M_{Z_\ell}$, where one obtains the correct relic density, i.e. $\Omega_\chi h^2=0.12$.
As one can appreciate, in this simple theory, one can satisfy the LZ experimental bounds~\cite{LZ:2022ufs} when $\sin \theta$ is below $0.01$, which is also in agreement with the experimental bounds of SM Higgs mixing angle,  $\sin {\theta}<0.3$. Therefore, one can easily satisfy the current experimental bounds from direct detection experiments.
\section{LEPTON NUMBER VIOLATING SIGNATURES AT THE LHC}
\label{LNV-LHC}
As we have discussed above, in this theory the seesaw scale can be below the ${\cal{O}}(10^2)$TeV scale in agreement with the cosmological bounds on the dark matter relic density, and one can hope to test the mechanism responsible for Majorana neutrino masses at colliders. In this article, we will focus on the testability at the Large Hadron Collider (LHC). The collider bounds on the leptophilic gauge boson, $Z_\ell$, are very strong and it will be very difficult to produce it at the LHC. However, the Standard Model-like Higgs  and new Higgses, $H_1$ and $H_2$, decays can give rise to lepton number violating signatures. For a previous study of lepton number violating signatures from Higgs decays see for example Refs.~\cite{Accomando:2016rpc,Deppisch:2018eth}.

One can produce the Higgses at the LHC through gluon fusion and look for the decays into right-handed neutrinos that give rise to lepton number violating signatures with two same-sign charged leptons and four jets:
\begin{equation}
pp \to h, H_i \to N_k N_k \to     e^{\pm}_j e^{\pm}_m 4 j.
\end{equation}
\begin{itemize}
    \item LNV Signatures from $h$ decays:
    The SM-like Higgs can decay into two right-handed neutrinos, if kinematically allowed, through the mixing between the Higgses. The decay width can be  written as
\begin{equation}
        \Gamma ( h \to N N)=  \frac{3}{4 \pi} \frac{M_N^2\hspace{0.1 cm}g_{\ell}^2\hspace{0.1 cm}\sin^2{\theta}}{M_{Z_{\ell}}^2 \cos^2{\beta}} M_h \left(1-\frac{4M_N^2}{M_h^2}\right)^{3/2},
\end{equation}
and the branching ratio is given by
\begin{equation}
    {\rm{BR}}(h \rightarrow NN)=\frac{\Gamma(h\rightarrow NN)}{\cos^2{\theta} \hspace{0.1 cm}\Gamma_{SM}+\Gamma(h\rightarrow NN)}.
\end{equation}
Here we are assuming for simplicity that all right-handed neutrinos have the same mass. In this case, we can estimate the number of events as given by
\beq
N_h (e^{\pm}_j e^{\pm}_m 4 j) = \mathcal{L} \times \sigma(p p \to h) \times {\rm BR} ( h \to N N ) \times 2 \times {\rm BR} (N \to e^{\pm}_j jj) \times {\rm BR} (N \to e^{\pm}_m jj).
\eeq
\begin{figure}[h]
     \centering 
     \begin{subfigure}[h]{0.5\textwidth}
         \centering
         \includegraphics[width=\textwidth]{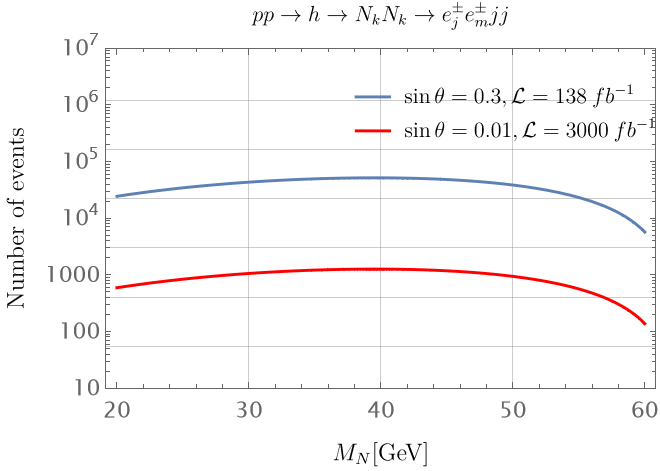}
     \end{subfigure}
    \caption{Lepton number violating events from the SM-like Higgs decays as a function of right-handed neutrino masses ($M_N$). The red line shows the predictions when the luminosity is
    $\mathcal{L}=3000  \hspace{0.1 cm}\text{fb}^{-1}$, 
  $\cos{\beta}=0.84$, BR($N_k \rightarrow e_i^{\pm}jj)=0.5 \hspace{0.2 cm}$ and $\sin \theta=0.01$. The blue line corresponds to the case when $\mathcal{L}=138  \hspace{0.1 cm}\text{fb}^{-1}$ and $\sin \theta=0.3$.}
    \label{LNV-SMHiggs}
\end{figure}
In Fig.~\ref{LNV-SMHiggs} we show the number of LNV events from the SM-like Higgs decays using the HL-LHC luminosity $\mathcal{L}=3000  \hspace{0.1 cm}\text{fb}^{-1}$, $\sigma (pp \to h)=54.7 \ \text{pb}$, $\cos{\beta}=0.84$ and $\sin{\theta}=0.01$. Notice that the mixing angle $\theta$ has to be small, $\sin{\theta} \leq 0.01$, to be in agreement with the dark matter direct detection constraints. The branching ratio ${\rm BR} ( h \to N N )$ is around $10^{-5}$ in the scenarios shown in Fig.~\ref{LNV-SMHiggs}. Since the dark matter direct detection bounds are only valid in a fraction of the parameter space, we show in Fig.~\ref{LNV-SMHiggs} the number of events when the luminosity $\mathcal{L}=138  \hspace{0.1 cm}\text{fb}^{-1}$, and $\sin{\theta}=0.3$.
This simple estimation tells us that the HL-LHC could have access to a large number of LNV signatures coming from the SM-like Higgs decays. These signatures can be quite exotic because the light right-handed neutrinos can be long-lived and then the LNV decays can give rise to displaced vertices.
Clearly, these signatures are quite unique but the predictions depends on the mixing angle $\theta$ between the Higgses that could be smaller. The main reducible background for these signals is $pp \to t \bar{t} W^{\pm} \to W^\pm W^\pm jj b \bar{b}$ but using a set of kinematic cuts one can reduce the background quite effectively. See for example the discussion in Ref.~\cite{FileviezPerez:2009hdc} for details.

    \item LNV Signatures from $H_i$ decays:
The new Higgses can be produced at the LHC though the mixing with the SM-like Higgs and their decays can give rise to LNV signatures.
In this case of number of events is given by
\begin{eqnarray}
N_{H}(e^{\pm}_j e^{\pm}_m 4 j) &=& \mathcal{L} \times \sigma(p p \to H_i) \times {\rm BR} ( H_i \to N_k N_k) \times 2 \nonumber \\ 
&\times & {\rm BR}(N_k \to e^\pm_j W^\mp) \times {\rm BR}(N_k \to e^\pm_m W^\mp) \times {\rm BR}^2(W^\mp\to jj),
\end{eqnarray}
where the hadronic decay of the $W$ boson is BR$(W^\mp\to jj)\simeq 2/3$. The right-handed neutrino decay width for $N_i \to e^-_j W^+$ is  given by
\begin{align}
\Gamma(N_i \to e^-_j W^+) & = {g^2_2 \over 64\pi
M_W^2}|V_{j i}|^2 M_{N_i}^3 \left(1+2\frac{M_W^2}{M_{N_i}^2}\right)  \left(1-\frac{M_W^2}{M_{N_i}^2}\right)^2. 
\label{eq:NlW} 
\end{align}
The matrix $V$ defining the mixing between the right-handed and left-handed neutrinos can be written 
as~\cite{Casas:2001sr},
\begin{eqnarray}
V = \ V_{\rm PMNS} \ m^{1/2} \ R \ M^{-1/2},
\label{mixing1}
\end{eqnarray}
where $V_{\rm PMNS}$ is the PMNS mixing matrix, $m={\rm{diag}} (m_1, m_2, m_3)$ is the matrix of the light neutrino masses 
and $M={\rm{diag}} (M_{N_1}, M_{N_2}, M_{N_3})$ is the matrix for the heavy neutrino masses, and the $R$ matrix is complex and orthogonal. For a detailed study of the ${\rm BR} (N_k \to e^\pm W^\mp)$ taking into account the neutrino constraints and using the freedom in the matrix $R$ see Ref.~\cite{FileviezPerez:2020cgn}.
 
\begin{figure}   
    \begin{subfigure}[h]{0.45\textwidth}
        \centering
         \includegraphics[width=\textwidth]{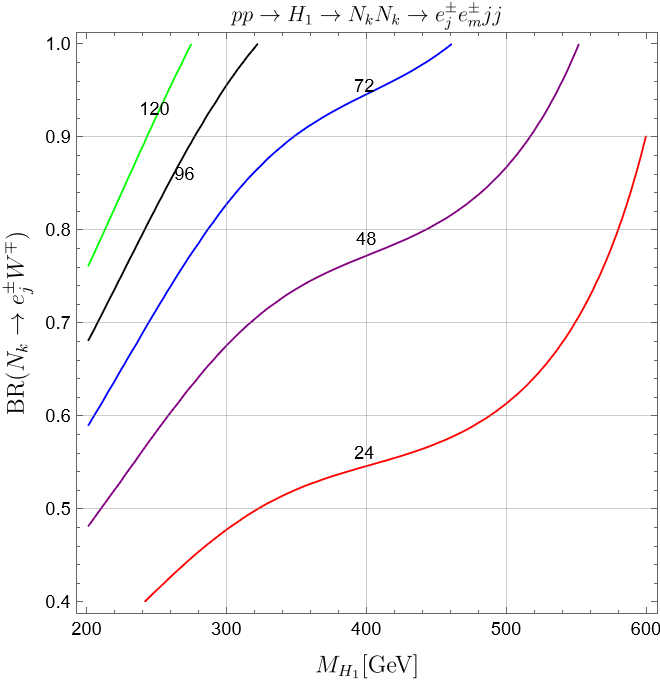}
     \end{subfigure}
     \hfill
     \begin{subfigure}[h]{0.45\textwidth}
         \centering
         \includegraphics[width=\textwidth]{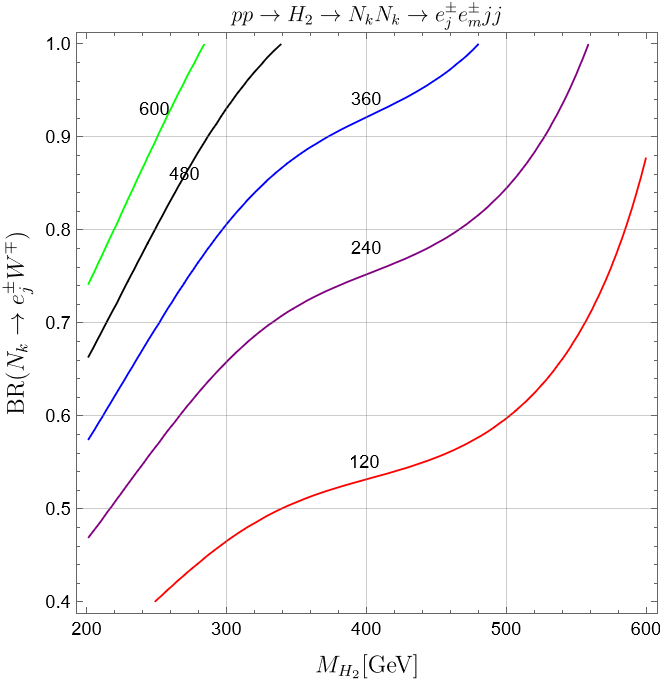}
     \end{subfigure}
  \caption{Number of lepton number violating events as a function of heavy Higgs  masses ($M_{H_i}$) and Br($N_k \rightarrow e^{\pm}_j W^{\mp}$). Here we used luminosity, $\mathcal{L}=3000  \hspace{0.1 cm}\text{fb}^{-1}$ and $\cos{\beta}=0.84$, 
  $M_N= M_{\chi}=100$ GeV, $\sin{\theta}=0.01$, $\sin{\theta_H}=0.5$, 
  and $M_{Z_{\ell}}/ g_\ell = 7$ TeV. }
    \label{LNV-NewHiggs}
\end{figure}
The cross-section $\sigma(p p \to H_i)=\sin^2 \theta \ \sigma(p p \to H_i)_{SM}$, where $\sigma(p p \to H_i)_{SM}$ is the predicted cross section in the Standard Model changing the Higgs mass. Since this cross section is suppressed by the mixing angle and $\sin \theta \leq 0.01$, to be in agreement with dark matter direct detection constraints, one cannot expect a large number of events even at the HL-LHC.

In the above equation, the branching ratio for the Higgs decays into two right-handed neutrinos is given by
\begin{equation}
    {\rm BR} ( H_i \to N_k N_k)=\frac{\Gamma ( H_i \to N_k N_k)}{\sin^2 \theta \ \Gamma (H_i \to {\rm SM} \ {\rm  SM}) + \Gamma (H_i \to \bar{\chi} \chi) + \Gamma ( H_i \to N_k N_k)+\Gamma ( H_i \to h h)},
\end{equation}
where $H_i \to {\rm SM} \ {\rm  SM}$ are the decays into SM particles excluding the SM-like Higgs. Since in our case $\sin \theta \leq 0.01$ these decays are very suppressed. The decay width for $H_i \to N_k N_k$ is given by
\begin{equation}
\Gamma (H_i \to N_k N_k) = \frac{3}{4 \pi} \frac{ M_N^2 g_{\ell}^2 U_{3i}^2 }{ M_{Z_{\ell}}^2 \cos^2{\beta}} M_{H_i} \left(1-\frac{4M_{N}^2}{M_{H_{i}}^2}\right)^{3/2},
\end{equation}
the decay width for  $H_i \to  \overline{\chi} \chi$ is 
\begin{equation}
\Gamma(H_{i}\rightarrow  \overline{\chi} \chi)=\frac{9}{8 \pi} \frac{M_{\chi}^2 g_{\ell}^2 U_{2i}^2}{M_{Z_{\ell}}^2 \sin^2{\beta}}
   M_{H_{i}}\left(1-\frac{4M_{\chi}^2}{M_{H_{i}}^2}\right)^{3/2},
\end{equation}
and the decay width for $H_i \to h h$ is 
\begin{equation}
    \Gamma(H_{i}\rightarrow hh)= \frac{\lambda_{H_ihh}^2}{32 \pi}\left(\frac{M_{Z_{\ell}}}{6 g_{\ell}}\right)^2 \frac{1}{M_{H_i}}\left(1-\frac{4M_h^2}{M_{H_{i}}^2}\right)^{1/2}
\end{equation}
where $$\lambda_{H_ihh} = 2 \lambda_1 \hspace{0.1 cm}  U_{2i} \sin{\beta}+3 \lambda_2 \hspace{0.1 cm} U_{3i} \cos{\beta}.$$
In Fig.~\ref{LNV-NewHiggs} we show the predictions for the LNV events at the HL-LHC as a function of heavy Higgs  mass, $M_{H_i}$, and Br($N_i \rightarrow e^{\pm}W^{\mp}$). For illustration, we use $\mathcal{L}=3000  \hspace{0.1 cm}\text{fb}^{-1}$, $\cos{\beta}=0.84$, $M_N= M_{\chi}=100$ GeV and $\sin{\theta}=0.01$. Here we neglected the decays into two SM-like Higgses to show the most optimistic scenarios. Unfortunately, the number of events from the heavy Higgses are not too large at the LHC due to the fact that the production cross-section is suppressed. Therefore, the best hope at the LHC is to look for the lepton number violating signatures from the SM-like Higgs decays into two right-handed neutrinos.

\end{itemize}
\newpage
\section{SUMMARY}
\label{Summary}
The origin of neutrino masses is one of the most pressing issues in particle physics.
In this article, we discussed a class of theories for Majorana neutrinos where the total lepton number is a local gauge symmetry. In order to define an anomaly free theory based on the total lepton number one needs to add extra fermions including the right-handed neutrinos. These theories predict a fermionic dark matter candidate from anomaly cancellation. The properties of the dark matter candidate were discussed in great detail taking into account all annihilation channels. The dark matter candidate has unique annihilation channels due to the fact that the theory predicts a very light pseudo-Goldstone boson, the Majoron, and one can obtain the correct relic density in a large fraction of the parameter space. Using the cosmological bounds on the dark matter relic density we pointed out the upper bound on the total lepton number symmetry breaking scale. 

These theories provide an unique scenario predicting a low scale seesaw mechanism for Majorana neutrino masses since the seesaw scale is below the ${\cal{O}}(10^2)$TeV scale. We have shown that one could hope to test the origin of neutrino masses at current or future colliders. We discussed the lepton number violating Higgs decays and the possibility to observe lepton number violation at the Large Hadron Collider. We have shown that the SM-like Higgs decays can provide an unique window to the origin of neutrino masses if the lepton number violating decays can be observed in the near future.

{\small {\textit{Acknowledgments:}}
The work of P.F.P. is supported by the U.S. Department of Energy, Office of Science, Office of High Energy
Physics, under Award Number DE-SC0024160.
This work made use of the High Performance Computing Resource in the Core Facility for Advanced Research Computing at Case Western Reserve University. }

\newpage
\appendix
\section{FEYNMAN RULES}
In our study, the dark matter candidate is the Dirac spinor, $\chi=\chi_L+\chi_R$. Neglecting the mixing between $\chi$ and the neutral components of the fermionic $SU(2)$ doublets, we list the simplified Feynman rules relevant to study the dark matter annihilation channels and direct detection cross-section:
\begin{eqnarray}
  \overline{\chi} \chi Z_{\ell}  & : & \hspace{0.5 cm} -i g_{\ell}\gamma^{\mu}(\ell_2 P_L+ \ell_1 P_R),  
  \\
  \overline{\chi} \chi H_i  & : & 
 \hspace{0.5 cm} i\frac{3g_{\ell}M_{\chi}}{M_{Z_{\ell}}} \frac{U_{2i}}{\sin{\beta}},
  \\
  \overline{\chi} \chi J  & : & \hspace{0.5 cm}i\frac{3g_{\ell}}{2M_{Z_{\ell}}}P_{\mu}^J\gamma^{\mu} \gamma^5, \\
 Z_{\mu}^{\ell} H_i J &:& \hspace{0.5 cm} i (2P_{\mu}^J)(3g_{\ell} \sin {\beta} U_{2i}+ 2g_{\ell} \cos {\beta}U_{3i}), 
 \\
 NNJ &:& \hspace{0.5 cm} i \frac{g_{\ell}}{ 2 M_{Z_{\ell}}}P_{\mu}^J \gamma^{\mu}\gamma^5,
 \\
  \nu \nu J &:& \hspace{0.5 cm} - i \frac{g_{\ell}}{ 2 M_{Z_{\ell}}}P_{\mu}^J \gamma^{\mu}\gamma^5,
  \\
 \nu \nu Z_{\ell}^{\mu} &:& \hspace{0.5 cm} i\frac{g_{\ell}}{2}\gamma^{\mu}\gamma^5,
 \\
 N N Z_{\ell}^{\mu} &:& \hspace{0.5 cm} -i\frac{g_{\ell}}{2}\gamma^{\mu}\gamma^5,
 \\
 \bar{e}eZ_{\ell}^{\mu}&:& \hspace{0.5 cm} -i g_{\ell}\gamma^{\mu}, 
 \\
 H_i NN &:& \hspace{0.5 cm}i\frac{ g_{\ell} M_N }{M_{Z_{\ell}} \cos{\beta} }U_{3i},
 \\
 H_i Z_{\ell} Z_{\ell} &:&  \hspace{0.5 cm }2i g_{\ell} M_{Z_{\ell}}\left( 2 U_{3i} \cos{\beta} + 3U_{2i} \sin{\beta}\right) g_{\mu \nu}.
\end{eqnarray}
We have used $v_s= \frac{1}{3}v\sin{\beta}$ and $v_\phi =\frac{1}{2} v \cos{\beta}$ to obtain the above Feynman rules.
Here we are neglecting the small mixing angle between the SM neutrinos and the right-handed neutrinos.
Notice that we are working in the basis where the $J$-interactions are invariant under the shift symmetry. We redefine the fields as follows:
\begin{eqnarray}
\nu_R &\to & \e^{-i \sigma_\phi/2v_\phi} \nu_R, \\
\e_R &\to & \e^{-i \sigma_\phi/2v_\phi} \e_R, \\
\ell_L &\to & \e^{-i \sigma_\phi/2v_\phi} \ell_L,\\
\chi_L &\to & \e^{i \sigma_s/2v_s} \chi_L,\\
\chi_R &\to & \e^{-i \sigma_s/2v_s} \chi_R.
\end{eqnarray}
Notice that the Majoron, defined as $J=\cos \beta \ \sigma_\phi + \sin \beta \ \sigma_s$, couples to fermions as a pseudo-Goldstone boson.

\newpage
\section{DARK MATTER ANNIHILATION CHANNELS}
\label{AppendixB}

The annihilation cross sections were calculated using FeynCalc~\cite{Shtabovenko_2016,Shtabovenko_2020,MERTIG1991345}.
Our dark matter candidate has the following annihilation channels: 
\begin{figure}[h] 
         \centering
         \includegraphics[width=1.1\textwidth]{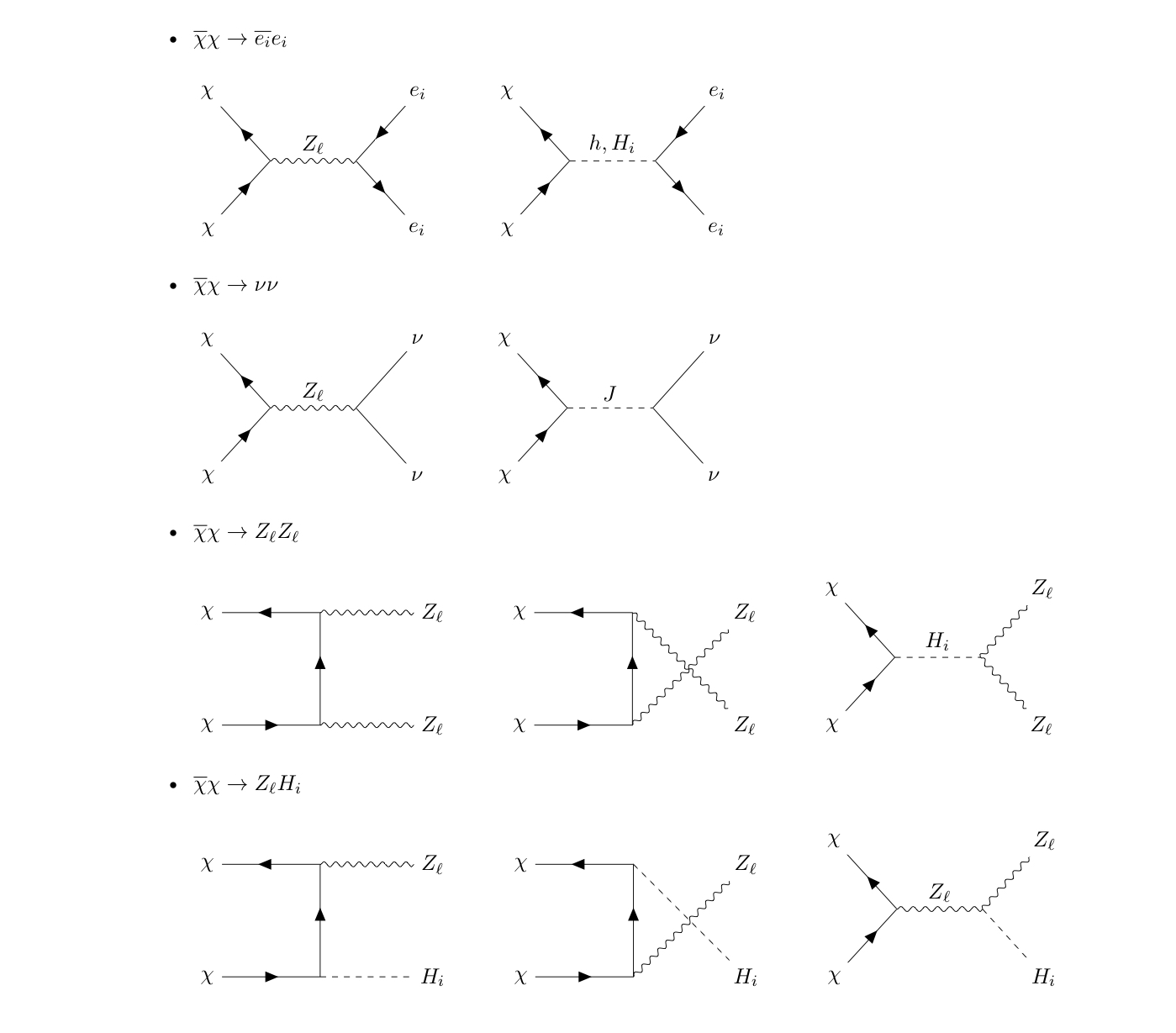}
\end{figure}
\newpage
\begin{figure}[h] 
         \centering
         \includegraphics[width=1.0\textwidth]{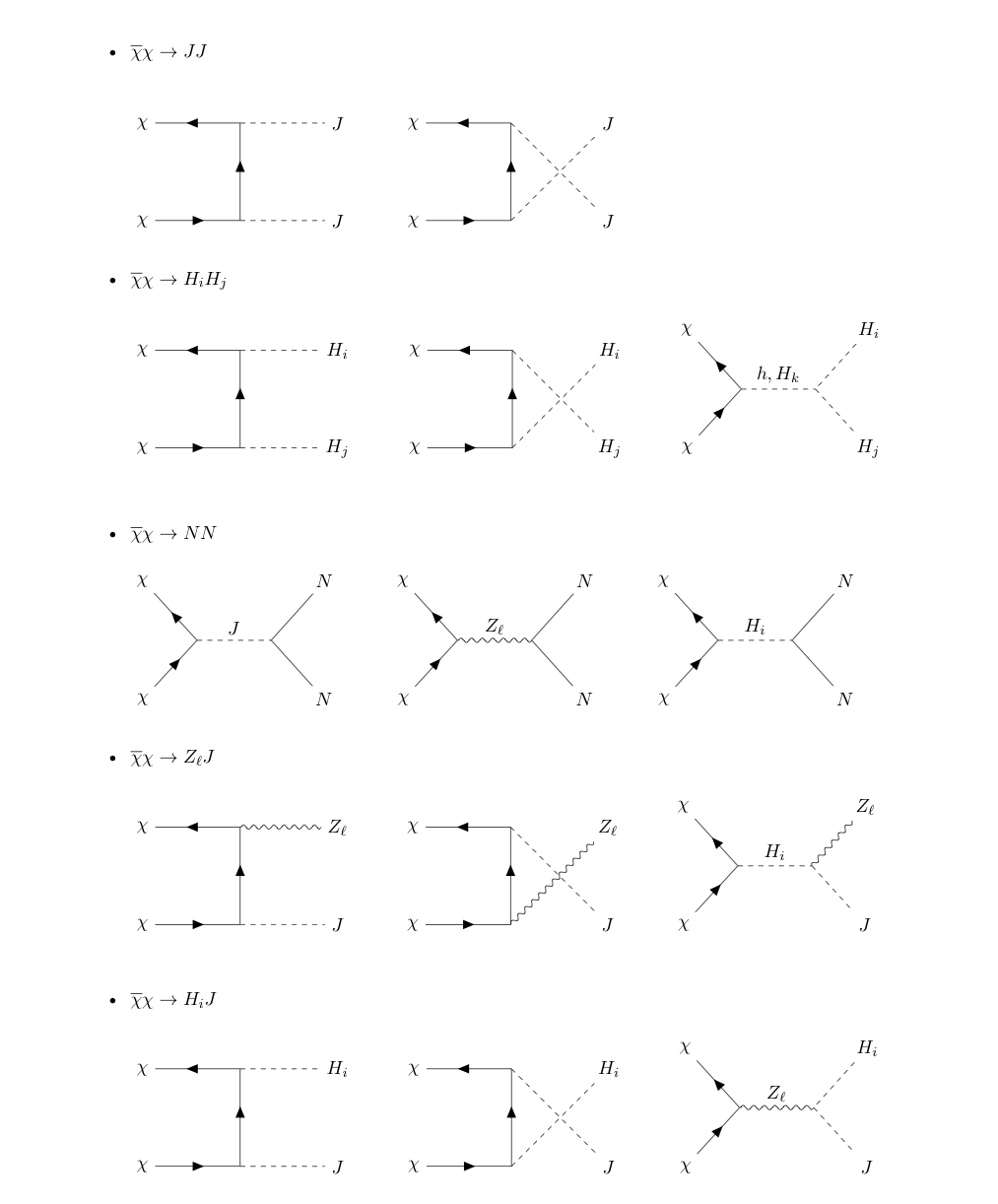}
\end{figure}
\newpage
\begin{figure}[h] 
         \centering
         \includegraphics[width=1.0\textwidth]{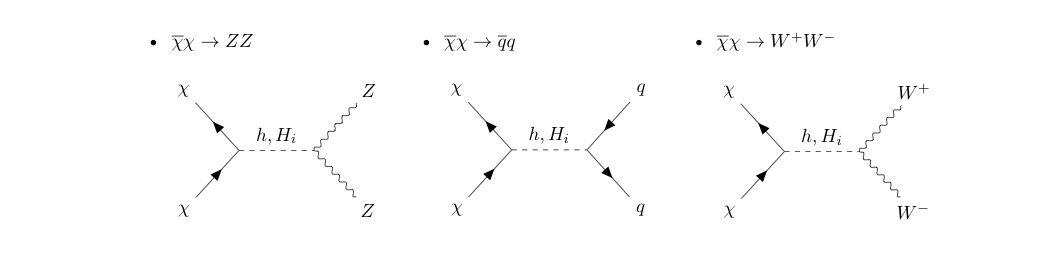}
\end{figure}

\bibliography{lepton}
\end{document}